\documentclass[%
 reprint,
 amsmath,amssymb,
 aps,
]{revtex4-2}

\usepackage{graphicx}
\usepackage{dcolumn}
\usepackage{bm}

\begin{document}

\preprint{APS/123-QED}

\title{Differential phase reconstruction of microcombs}

\author{Krishna Twayana$^{1}$}
\email{twayana@chalmers.se}
\author{ Fuchuan Lei$^{1}$}
\author{ Zhichao Ye$^{1}$}
\author{Israel Rebolledo-Salgado$^{1,2}$}
\author{Óskar B. Helgason$^{1}$}
\author{Magnus Karlsson$^{1}$}
\author{Victor Torres-Company$^{1}$}


\altaffiliation{$^1$ Department of Microtechnology and Nanoscience (MC2), Photonics Laboratory Chalmers University of Technology, SE-41296, Sweden \\}
\altaffiliation{$^2$ Measurement Science and Technology, RISE Research Institutes of Sweden, SE-501 15 Borås, Sweden}

\date{\today}

\begin{abstract}
Measuring microcombs in amplitude and phase provides unique insight into the nonlinear cavity dynamics but spectral phase measurements are experimentally challenging. Here, we report a linear heterodyne technique assisted by electro-optic downconversion that enables differential phase measurement of such spectra with unprecedented sensitivity (-50 dBm) and bandwidth coverage ($>$ 110 nm in the telecommunications range). We validate the technique with a series of measurements, including single cavity and photonic molecule microcombs. 
\end{abstract}

\maketitle


Microresonator-based Kerr frequency combs (microcombs) have attracted great attention as an optical source and witnessed significant research progress in the last decade. They enable applications ranging from spectroscopy to optical frequency synthesis \cite{pasquazi2018micro}. Stable microcombs can be generated in various forms including dissipative solitons \cite{herr2014temporal}, Turing rolls \cite{qi2019dissipative}, dark-pulses \cite{xue2015mode} or soliton crystals \cite{cole2017soliton}. In addition, the cavity can be engineered to display higher order dispersion, resulting in the emission of dispersive waves \cite{brasch2016photonic}; and multiple cavities can be linearly coupled to each other, resulting in unique coherent states \cite{helgason2021dissipative,tikan2021emergent}. Investigating the complex temporal shape of these waveform provides insight into the nonlinear dynamics but the measurement using conventional techniques \cite{trebino1997measuring} such as frequency-resolved optical gating or spectral phase interferometry for direct electric-field reconstruction is challenging. The pulse waveforms coming out of the cavity have usually very low energy due to the inherently large repetition rate. In addition, some microcomb waveforms are not transform-limited \cite{xue2015mode,helgason2021dissipative}, which reduces the efficiency for nonlinear gating.

A common strategy implemented in the context of microcomb spectral phase measurements is to equalize the phase of the microcomb with a pulse shaper until a transform-limited pulse is inferred from an intensity autocorrelation measurement \cite{ferdous2011spectral,del2015phase}. This technique played an instrumental role in unraveling the chaotic dynamics in microcombs \cite{ferdous2011spectral} and the discovery of dark-pulse Kerr combs \cite{xue2015mode} and soliton crystals \cite{cole2017soliton}. Background-free intensity autocorrelation measurements with high repetition rate pulses require an optical amplifier to boost the signal power and induce sufficient second harmonic power in the intensity autocorrelator. As a result, the spectral phase measurement is limited to the gain bandwidth of the amplifier or the bandwidth of the pulse shaper, whichever is narrower. An alternative diagnostic tool using dual comb interferometry was reported in \cite{kong2019characterizing,yi2018imaging}. This is a linear technique, first proposed in the context of frequency comb complex measurements in \cite{ferdous2009dual}, which can also enable extremely fast acquisition rates \cite{duran2015ultrafast}. Dual-comb interferometry allowed for retrieving experimentally the pathway to soliton formation. One drawback of this technique is that it relies on electric-field cross-correlation and thus requires a well-calibrated reference comb.

A linear broadband technique named stepped heterodyne has been successfully implemented for measuring the phase difference between consecutive lines in electro-optic frequency combs \cite{reid2010stepped} and modelocked lasers \cite{murdoch2011spectral, moskalenko2014record}. The idea lies in beating the signal waveform with a continuous-wave laser that is tuned in a stepped-wise manner across the signal comb lines. The phase of the consecutive lines is embedded in the downconverted radio-frequency beat notes \cite{reid2010stepped}. Here, we adapt this technique for the complex measurement of microcombs. This represents a great challenge as the repetition rate easily exceeds the bandwidth of state of the art photodetectors. We circumvent this issue by combining the technique with electro-optic down-conversion \cite{szafraniec2004swept}.
\begin{figure}[htb]
\centering
\includegraphics[width=\linewidth]{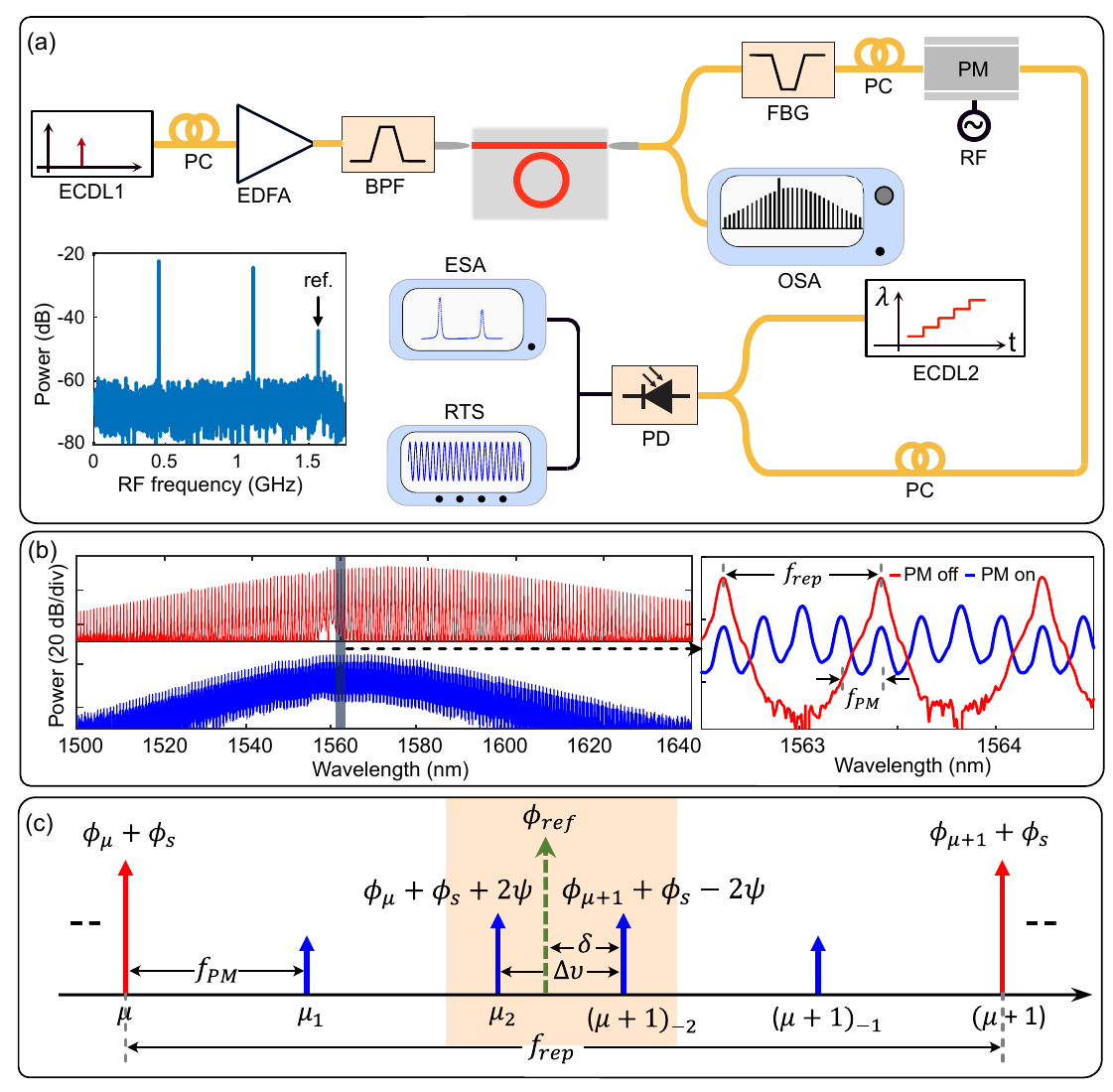}\caption{(a) Experimental setup for wideband differential phase measurement of microcombs based on electro-optic modulated stepped heterodyne spectroscopy. The inset shows the RF beating spectrum when the reference laser is positioned at 1557.2395 nm. The third peak acts as a reference signal for the phase difference calculation between the pair of comb modes. The frequency comb power was stabilized via a PID controller acting on the frequency of the pump laser (not shown). A tunable fiber Bragg grating (FBG) notch filter was used to suppress the pump power which otherwise saturates the photo-detector. (b) Single soliton comb (red) and electro-optic phase modulated comb (blue). Zoom (right) into both spectra. The spacing of the second-order sidebands is $\Delta\nu<1.6$ GHz and cannot be distinguished by the OSA resolution. The power of sidebands is related to the Bessel function. (c) Schematic spectrum of the phase modulated comb modes  $\mu$ and $\mu+1$ with a reference laser (green) positioned between the second-order sidebands. The phase of different comb modes is labelled. The highlighted lines at the center contribute to the three distinct downconverted beatnotes shown in the inset in (a).}
\label{fig:schematic}
\end{figure}
The experimental setup for the measurement of the spectral amplitude and phase is illustrated in Fig.\ref{fig:schematic}(a), where the former is directly measured on an optical spectrum analyzer (OSA). Basically, the differential phase is recovered by subtracting the phase of beating signals between the reference laser (ECDL2) and the sidebands of the comb lines. We used an electro-optic phase modulator (PM) to generate the sidebands (Fig. \ref{fig:schematic}(b) blue spectrum) that enables direct referencing of microcomb line spacing to a microwave frequency. The RF modulation frequency ($f_{PM}$) is varied in order to adjust the closest sideband spacing at the center within the detectable range. The phase modulation of the comb generates harmonics at $f_\mu + nf_{PM};(n\in\mathbb{Z})$ where, $\mu$ represents the microcomb mode number and $n$ the electro-optic sideband number. The modulated microcomb field can be represented as $\text{Re}(\sum_{\mu}\sum_{n}J_n(\beta)\text{exp}(j(2\pi f_\mu t+n(2\pi f_{PM}t+\psi)+\phi_\mu+\phi_s)))$. Here, $f_\mu$ is the frequency of $\mu^{th}$ comb line with static phase $\phi_{\mu}$ and $J_{-n}(\beta)=(-1)^n J_n(\beta)$ is the $n^{th}$ order Bessel function of the first kind at modulation index $\beta$. The phase of the RF signal is $\psi$ and phase noise of the optical carrier of the frequency comb is $\phi_s(t)$. The sidebands preserve the phase signature of the comb lines with an additional innocuous RF phase offset $n\psi$. Figure \ref{fig:schematic}(c) is the schematic of two comb lines and corresponding sidebands. The phase associated with the comb modes, sideband, and reference laser (green) are labelled. It is assumed that the optical phase noise of the comb is correlated among the comb lines. The reference laser is detuned by $\delta$ frequency from the second-order sideband of $\mu+1$ mode. The static phase of the mode $\mu$ is $\phi_{\mu}$ and for $\mu+1$ is $\phi_{\mu+1}$. The phase noise of the reference laser is $\phi_{ref}(t)$. The phase noise terms ($\phi_s, \phi_{ref}$) vanish from the differential phase as these terms are common to both the beating signals. With the electro-optic downconversion, the microcomb mode spacing $f_{rep}$ can directly relate to the $f_{PM}$ and a beat note ($\Delta\nu$) of the n\textsuperscript{th} order sidebands between the adjacent microcomb modes by $f_{rep}=n \times f_{PM}\pm \Delta\nu$ \cite{del2012hybrid}.  
\begin{figure*}[htb]
\centering
\includegraphics[width=\linewidth]{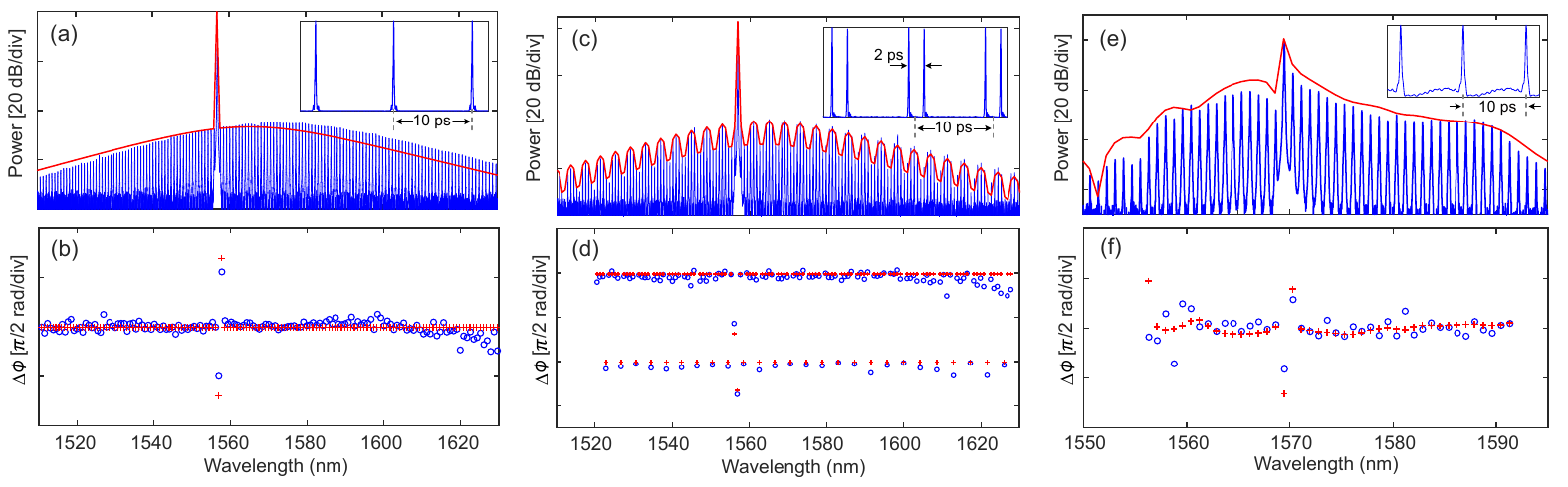}\caption{Spectral and temporal (inset) characterization of the microcombs (red simulation and blue measurement). The pulse intensity profiles in the insets are the recovered waveforms coupled out of the cavity including the pump mode (pump power is attenuated by the FBG in (a) and (b) insets). The measured differential phase corresponds to the spectrum inside the cavity except at the pump frequency. Spectral intensity profile: (a) single soliton comb ($f_{rep}$ 99.8867 GHz) (b) multisoliton comb ($f_{rep}$ 99.9011 GHz). The discrepancy in the repetition rate of the combs from the same device is attributed to the different pump detuning and coupling power variation \cite{zhang2019terahertz}. (c) coupled cavity comb ($f_{rep}$ 100.799 GHz). Measured spectral phase difference after subtracting the setup dispersion: (d) single soliton comb, (e) multisoliton comb, (f) coupled cavity comb. The phase offset at the pump frequency is a combined effect of the self-organization in the soliton formation \cite{wen2016self} and superimposed pump field at the through port \cite{wang2016intracavity}. The phase contribution of the latter interaction can be compensated from the comb spectral amplitude as reported in \cite{xue2015mode} and thus recovered the accurate intracavity time-domain waveform or simply by extracting the spectrum from the drop-port \cite{wang2016intracavity}.}
\label{fig:singlesolitoncomb}
\end{figure*}

In the measurement, the second-order sidebands of the stable frequency comb are heterodyned with the reference laser. The first two RF peaks in Fig. \ref{fig:schematic}(a) illustrate the resulting beat notes observed on an electrical spectrum analyzer (ESA). As the differential phase measurement is done on a pair of comb lines at a time, the reference laser is stepped at the $f_{rep}$ of the comb. The reference laser position is not critical, as long as it is within the accuracy  of $\Delta\nu/2$. The optical mixing translates the phase difference between the adjacent optical spectrum to an easily measurable low-frequency RF signal. The beat signals have power proportional to the product of the reference laser and the sideband power. As such, the high-power reference laser allows resolving the beat notes of weak comb line powers without resorting to an optical amplifier. 

In the differential phase calculation, an RF signal of frequency $\Delta\nu$ is required as reference. It is obtained from the beating of the nearest 2\textsuperscript{nd} sidebands in the photodetector (Fig. \ref{fig:schematic}(a) third peak in the ESA plot). However, it requires the optical signal to possess an intensity modulation (IM) of envelope frequency $\Delta\nu$. This is not the case with the phase modulated signal. Therefore, we sent a fraction of modulated spectrum to an optical programmable filter (OPF) used as a periodic narrow bandpass filter centred in between the comb lines (not shown in the schematic). 

A real-time scope (RTS) was used for the acquisition of the heterodyne signal detected by a photodiode of bandwidth (1.6 GHz) $>\Delta\nu$. The RTS acquires a ten microsecond interval of a beating signal. We record the RTS signal each time the reference laser is stepped to the next comb lines pair. The Fourier processing of the recorded signal results in three non-trivial RF components at $\Delta\nu-\delta$, $\delta$, and $\Delta\nu$. The product of first two components generates an RF signal of frequency $\Delta\nu$ whose phase is $\phi_{\mu+1}-\phi_\mu-4\psi+\psi_{res}$, with complete cancellation of the phase noise of both the comb source and reference laser. The synthesized RF signal is multiplied with the conjugate of extracted $\Delta\nu$ signal. This enables the extraction of the phase difference $\phi_{\mu+1}-\phi_\mu-4\psi-\phi_{net}+\psi_{res}$. Here, the constant phase offset $\phi_{net}$ from the reference signal is attributed to the beating of all the second order sidebands passing through the OPF. The phase offsets $\phi_{net}$ and $4\psi$ merely results in a linear phase term and temporal shift in a reconstructed signal. However, the measured differential phase includes both the inherent comb phase profile and the accumulated dispersion of the fiber assemblies at the output port of the microresonator. We calibrated this residual dispersive phase ($\psi_{res}$) with the aid of a swept-wavelength interferometer \cite{twayana2021frequency}. The device under test includes the components immediately outside the chip up to the input to the PM. Finally, the phase difference of the comb lines with a constant offset was calculated by subtracting the measured differential phase with the phase difference of the assemblies. Integration of the phase differences $\Delta\phi$ results in the relative phase of each comb line upon an otherwise irrelevant linear phase term. This causes a temporal shift when reconstructing the pulse. The intensity profile (Fig. \ref{fig:singlesolitoncomb} insets) was inferred from the complex optical spectrum as $|{\sum_{\mu}\sqrt{P_{\mu}}e^{i\mu 2\pi f_{rep}t+i\phi_{\mu}}}|^2$ where, $P_{\mu}$ is power and $\phi_{\mu}$ is the reconstructed phase of the comb mode $\mu$.

A crucial aspect of this technique is that it can accurately retrieve the phase profile at extremely low power levels. Remarkably, there is no optical amplifier in the heterodyne setup, allowing for measuring extremely broad microcombs. These features are exemplified with microcombs generated in two different low loss silicon nitride ($\text{Si}_3\text{N}_4$) microresonator configurations: a single ring with anamolous dispersion and coupled cavity rings with normal dispersion. The microresonators were fabricated via a subtractive processing method reported in \cite{ye2019high}. The frequency comb was measured at the through port for both cases. We consider a soliton microcomb generation in the single cavity microresonator of free spectral range (FSR) $\sim$100 GHz. The same device was used to generate a soliton crystal and multi-soliton consisting of two circulating solitons. The frequency comb is initialized by thermal kicking technique \cite{joshi2016thermally} and proceeds to a single soliton state through a slow forward frequency tuning \cite{guo2017universal}. To generate the comb in the coupled cavity, an auxiliary resonance is first slightly blue-detuned from the main cavity resonance by applying a voltage to a heater circuit. The microcomb is then deterministically generated by blue tuning the pump laser into the main cavity resonance \cite{helgason2021dissipative}.  

Numerical simulations of the different comb states were conducted with the Ikeda map to analyze the spectrum and differential phase profiles (Fig. \ref{fig:singlesolitoncomb} red profiles). For the simulation, some parameters were measured and have the following values: extrinsic coupling rate $\kappa_{ex}=14\times10^6$ Hz, intrinsic loss rate $\kappa_{in}=13\times10^6$ Hz, and dispersion $\beta_2=-65$ ps$^2$km$^{-1}$.  For the photonic molecule frequency comb simulation, we followed ref. \cite{helgason2021dissipative}, where the Raman term is turned off because it plays a negligible effect. The FSRs for the two rings are 100.8 GHz and 99.83 GHz respectively. The main cavity has internal loss rate $\kappa_{in}=38\times10^6$ Hz and coupling rate $\kappa_{ex}=32\times10^6$ Hz. The mode coupling induced mode splitting is 607 MHz and dispersion 100 ps$^2$km$^{-1}$.  

The measurement results (blue) validated with the simulations for three different microcomb states are illustrated in Fig. \ref{fig:singlesolitoncomb}. The spectrum of the single FSR soliton comb at a pump wavelength 1557.147 nm with on-chip power $\sim20.5$ dBm is shown in Fig. \ref{fig:singlesolitoncomb}(a). The power at the blue side edge of the spectrum is $< -50$ dBm. The spectrum is blue-detuned by $\sim$1.7 THz  relative to the soliton spectral maximum due to the Raman self-frequency shift \cite{yi2016theory}. The RF clock to the PM was set to 24.58 GHz, yielding $\Delta\nu=1.5667$ GHz spacing between second-order sideband from the adjacent modes. The spectral phase difference after correcting the setup dispersion is shown in Fig. \ref{fig:singlesolitoncomb}(b). The phase difference at the pump location is shifted approximately by 0.5$\pi$. 

Figure \ref{fig:singlesolitoncomb}(c) shows the simulated (red) and measured (blue) spectral profile of a multisoliton comb. The frequency comb has a repetition rate of 99.9011 GHz. It has a spectral modulation where one comb line in every five becomes attenuated. This 5 FSR modulation in the comb envelope indicates two intracavity pulses of relative time difference $\sim$2 ps (Fig. \ref{fig:singlesolitoncomb}(c) inset). The $\Delta\phi$ of the comb line pairs are aligned, except for the weaker comb lines which have a relative phase difference of $\pi$ (Fig. \ref{fig:singlesolitoncomb}(d)). At the pump location, it has also $\Delta\phi\sim0.5\pi$ as in the single cavity microcomb. The discrepancy between simulation (red) and measurement (blue) is likely due to uncertainties in the Raman coefficient in silicon nitride and the fact that our simulations do not include a wavelength dependent coupling coefficient. The phase difference is deviating from the simulated trend beyond 1600 nm wavelength (Fig. \ref{fig:singlesolitoncomb} (b,d)). It may be related to the wavelength dependent power coupling.

\begin{figure}[htbp]
\centering
\includegraphics[width=0.9\linewidth]{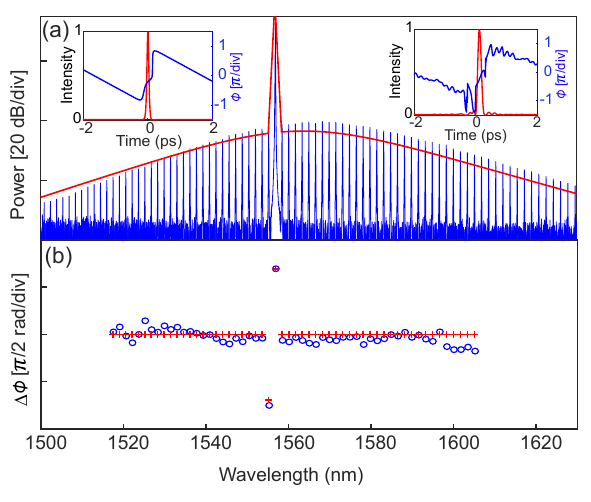}\caption{Complex spectral and temporal characterization of a 2-FSR soliton microcomb. The modulation depth of the PM was adjusted to bring the fourth-order sidebands within the range of the photodetector. The pump laser source is located at 1557.180 nm. (a) Spectrum of the microcomb of a repetition rate 199.7843 GHz. Insets are the reconstructed temporal intensity and phase profiles (left simulation and right measurement). (b) Phase difference measurement of the comb modes.}
\label{fig:single2FSRsolitoncomb}
\end{figure}

The microcomb coming out of the photonic molecule is shown in Fig. \ref{fig:singlesolitoncomb}(e). The comb spectrum was acquired from the power coupled out of the bus waveguide. The cavity was pumped at 1569.67 nm with an on-chip optical power of 13 dBm. The microcomb has a repetition rate of 100.799 GHz. Unlike the flat $\Delta\phi$ in the single cavity microcomb, the coupled cavity microcomb has non-uniform $\Delta\phi$ (Fig. \ref{fig:singlesolitoncomb} (f)). The equivalent pulse coupled out of the cavity is shown in the inset of Fig. \ref{fig:singlesolitoncomb}(e). The waveform of the coupled cavity microcomb is not transfer-limited. Indeed, it is a chirped pulse with asymmetric damping oscillation at both sides of the pulse (Fig. \ref{fig:singlesolitoncomb}(e) inset) in line with the measurements and predictions of \cite{helgason2021dissipative}.

The proposed technique is in principle not limited by the repetition rate of the comb. Figure \ref{fig:single2FSRsolitoncomb} demonstrates the characterization of a 2-FSR comb having 199.7843 GHz comb spacing. The RF frequency and modulation depth of the PM was adjusted such that the sidebands at the center are stronger and within the detectable frequency range. In this measurement, we increased the 24.78 GHz RF power applied to the PM. This in turn generates higher-order sidebands. For the heterodyning, we optimized the 4\textsuperscript{th} order sideband from the neighbouring comb modes. Figure \ref{fig:single2FSRsolitoncomb}(b) is the calculated differential phase of the 2-FSR comb. The accuracy of the measured differential phase values is estimated to 0.2 rad, and the resulting phase error can thus be expected to grow as 0.2 times the square root of the number of summed differential phases. The reconstructed temporal intensity and phase profiles in Fig. \ref{fig:single2FSRsolitoncomb}(a) insets shows good agreement between simulation (left) and measurement (right).  

In summary, we have demonstrated broadband complex spectral characterization of microcombs using stepped heterodyne interferometry combined with electro-optic downconversion. It requires no optical amplifier to boost the comb lines power, which allows for retrieving comb states over a bandwidth $>110$ nm, only limited by the tuning range of the stepped lasers. The technique is linear and can also measure the phase difference of extremely weak power lines. We have validated the technique with a number of microcomb states produced in silicon nitride microresonators.

This project is financially supported by the Horizon 2020 Framework Programme (GA 812818), the Swedish Foundation for Strategic Research (FID16-0011), the European Research Council (GA 771410 DarkComb), and Vetenskapsrådet (2016-03960, 2016-06077, 2020-00453).
\nocite{*}

\bibliography{apssamp}

\end{document}